# Investigation of the Time Performance of a LYSO Array for TOF-PET

Liu Jun-Hui(刘军辉)[1,2], Xu Jiong-Hui(徐炯辉)[3], Cheng Feng-Feng(成枫锋)[4,5], Li Dao-Wu(李道武)[1,2], Zhang Zhi-Ming(章志明)[1,2], Wang Bao-Yi(王宝义)[1,2], Wei Long(魏龙)[1,2]*

(1.Institute of High Energy Physics, Beijing 100049, China；
2. Beijing Engineering Research Center of Radiographic Techniques and Equipment；
3. Qinshan Nuclear Power Station, Haiyan, 314300, China；
4. Key Lab of Beam Technology and Material Modification of Ministry of Education, College of Nuclear Science and Technology，Beijing Normal University，Beijing 100875，China
5. Beijing Radiation Center, Beijing 100875, China)

**Abstract:** Positron Emission Tomography (PET) using time-of-flight information, which can greatly improve the quality of the reconstructed image, has recently become an exciting topic. In this work, 3.2mm×3.2mm×25mm finger-like LYSO crystals were used to make a 5×5 array, coupled to the Hamamatsu H8500 photomultiplier tube (PMT) as a detector. A fast leading-edge discriminator was designed for the LYSO-H8500 detector. Average coincidence time resolution FWHM of 330 ps was obtained for the LYSO detector with a reference $BaF_2$ detector, whose time resolution for 511 keV γ-rays was FWHM 150 ps. Time resolution FWHM of 294 ps was calculated for the LYSO detector, and coincidence time resolution of FHWM 415 ps can be expected for two identical LYSO detectors.

## 1. Introduction

The application of time-of-flight detector technology to Positron Emission Tomography allows an obvious increase in image contrast, reduction of the dose and shorter scanning time, which has attracted many people's attention [1-3]. Position sensitive photomultiplier tubes (PS-PMTs) not only have excellent spatial resolution, but also have very fast rise time due to the small gaps between the dynodes, and some even have small transit time spread comparable to other fast PMTs, which makes them good for timing applications [4]. LYSO (or LSO) scintillator has proved to be a good candidate for TOF-PET due to its high light output, fast decay time and strong stopping power [5-6]. PS-PMTs and LYSO scintillator together could therefore be a good choice for a TOF-PET detector. T. Moriya designed a TOF-PET detector using 2.9mm×2.9mm×20mm LYSO crystals coupled to a Hamamatsu R8400-00-M64 PMT and got FWHM 505 ps coincidence time resolution with a $BaF_2$ reference detector [7]. Chang Lyong Kim designed a TOF-PET detector using 4.2mm×4.2mm×30mm LYSO crystals coupled to a Hamamatsu H8500 PMT and got FWHM 477 ps coincidence time resolution (crystal to crystal)[4] .

Timing discrimination is an important factor influencing the performance of TOF-PET detectors. Normal timing discriminators like the constant fraction discriminator (CFD) ORTEC 935 are designed for negative signals. However, it is more convenient for PS-PMTs to use positive signals from the dynode for timing, because the signals share the dynodes and go to separate anodes, so we can consider using negative high voltage for the PMT. Besides, a leading-edge (LE) discriminator might be more suitable than CFD for LYSO scintillator [8]. So, it is useful to design a fast LE discriminator for positive signals for TOF-PET detectors based on the LYSO scintillator and PS-PMT.

In this paper, a 5×5 LYSO array coupled to a Hamamatsu H8500 PMT was used for the TOF-PET detector and a fast LE discriminator was designed for the detector. The performance of the detector was tested correctly. This could be useful for people studying TOF-PET detectors.

## 2. EXPERIMENTAL APPARATUS
## 2.1 LYSO-H8500 DETECTOR

As shown in Figure 1, the LYSO-H8500 detector was composed of a 5×5 array of 3.2mm×3.2mm×25mm LYSO crystals and the H8500 PMT. The surfaces were polished. There was nothing between the pixels, which means air was the only medium to help decrease the light crosstalk by total reflection. From experience, the crosstalk caused by air-gap would slightly affect the spatial resolution. However, the scatter map was clear enough, as can be seen in Figure 2. The time resolution would also be slightly affected by the crosstalk caused by the air-gap. However, this is compensated for by the better light collection which arises from this structure. Teflon was wrapped around the array to reflect light, and the top of the array was also wrapped in Teflon. The array was optically coupled to the PMT with silicon grease, which does not go into the gap between the pixels and is better than silicon oil because of the strong adhesion. The scatter map is shown in Figure 2.

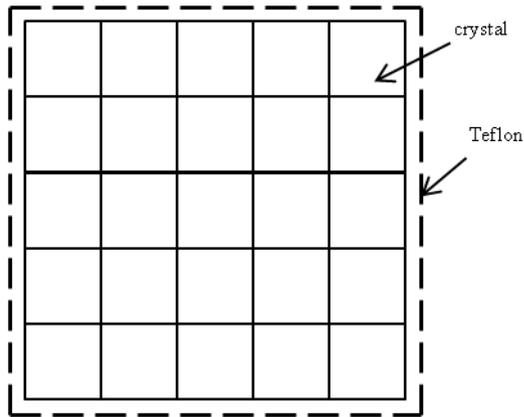

Figure 1 Sketch map for the LYSO-H8500 detector

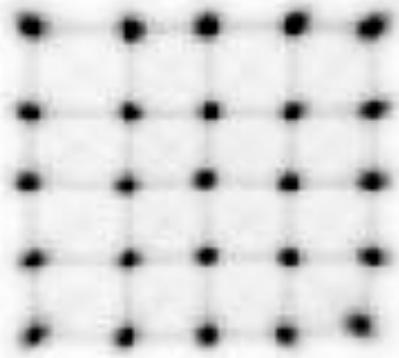

Figure 2 Scatter map of the LYSO-H8500 detector

## 2.2 Leading-edge discriminator

The schematic of the LE discriminator is shown in Figure 3 and Figure 4. The dynode signal of the PMT was split and sent to Comparator 1 and Comparator 2. The former was used to time the dynode signal and the latter was used to suppress noise signals. The output of Comparator 1 was delayed to trigger the CLK of the flip-flop. The output of Comparator 2 went into the D of the flip-flop to open the gate for the signal of Comparator 1. The output of the flip-flop was shaped as a constant width signal by the shaping circuit.

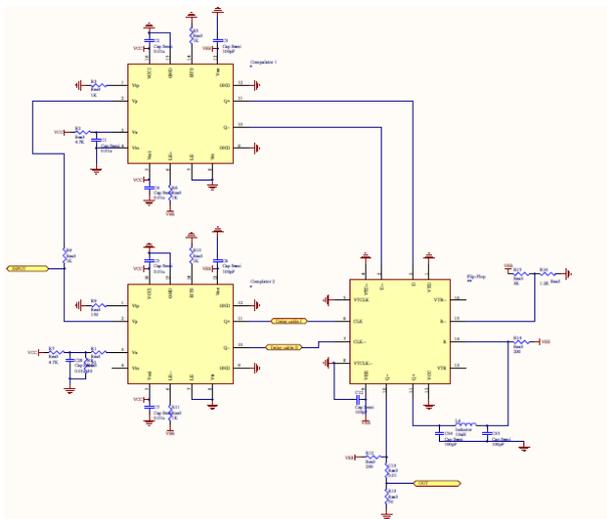

Figure 3 Schematic of the LE discriminator

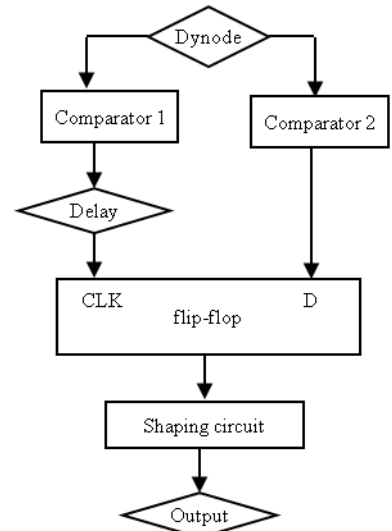

Figure 4 Diagram for the LE discriminator

## 2.3 BaF$_2$-XP2020Q detector

The BaF$_2$-XP2020Q detector was composed of a Φ30mm×20mm crystal and a XP2020Q PMT. The time resolution of the detector was 150 ps, measured with the help of two other BaF$_2$ detectors.

If the time resolutions of the 3 BaF$_2$ detectors are respectively X, Y and Z,
then
$$X^2+Z^2=(221ps)^2$$
$$Y^2+Z^2=(229.3ps)^2$$
$$X^2+Y^2=(221ps)^2$$
$$X=150ps$$

## 2.4 The test system

Test system (a) is a typical fast-slow coincidence test system, with the LYSO-H8500 detector as the start and the BaF$_2$-XP2020Q detector as the stop, as shown in Figure 5. The timing information of the LYSO-H8500 detector is extracted from the inverted dynode signal, triggering the CFD ORTEC 935.

Figure 6 shows test system (b), where the timing information of the LYSO-H8500 detector is extracted from the dynode signal, triggering the LE discriminator.

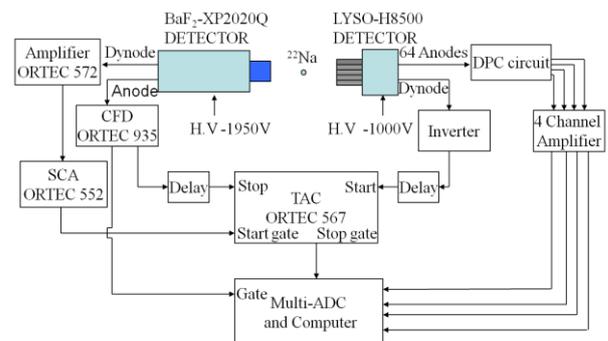

Figure 5 Fast-slow coincidence system (a)

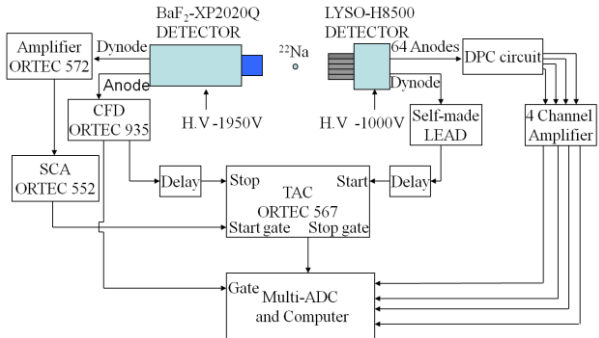

Figure 6 Fast-slow coincidence system (b)

3.  Test result

The light output of the 3.2mm ×3.2mm ×25mm LYSO crystal was measured as 3226 p.e/MeV by the single photon method [9], wrapped with Teflon and standing on the H8500 PMT, coupled with silicon oil.

The light pulse curve of the LYSO scintillator was also measured with the single photon method [9], shown in Figure.7. The rise time of the curve was computed to be 1.8 ns, and the decay constant was 42.4 ns.

By comparison, the dynode signal of the LYSO-H8500 detector was measured with a 500M oscilloscope, shown in Figure 8. The rise time of the signal was 1.7 ns.

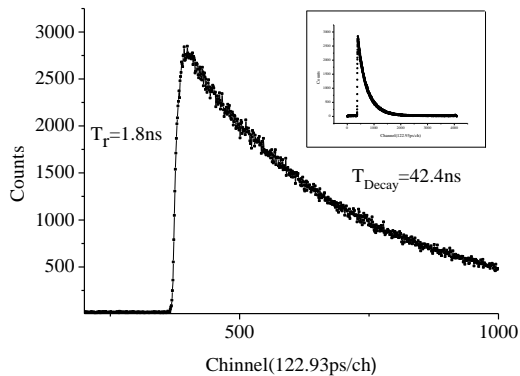

Figure.7 Light pulse curve of the LYSO crystal

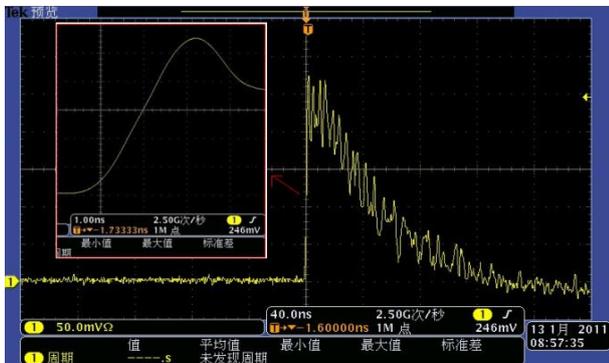

Figure.8 The large inset shows the dynode signal for the LYSO-H8500 detector for 511keV; the small inset shows the leading part of the magnified signal.

The coincidence time spectrum of the center LYSO crystal in the array and the BaF$_2$ detector were measured respectively with test system (a), shown in Figure 9, and test system (b), shown in Figure 10. The calculated time resolution is shown in Table 1.

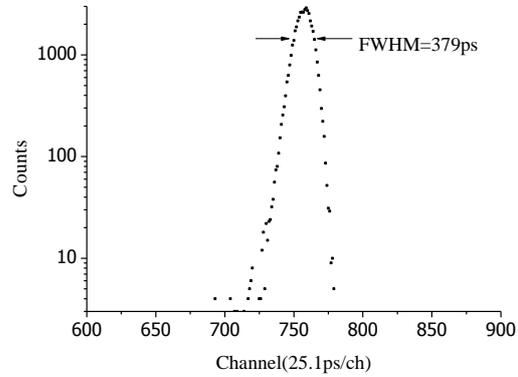

Figure.9 Coincidence time spectrum tested with system (a)

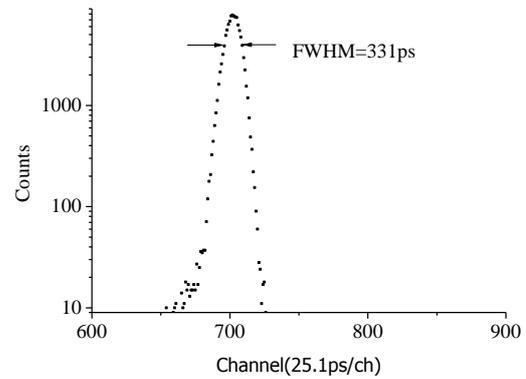

Figure.10 Coincidence time spectrum tested with system (b)

Table.1 Detector timing resolution

| Timing setup | TR$_S$(ps) | TR$_L$（ps） | TR（ps） |
|---|---|---|---|
| System (a) [Inverter+ CFD ORTEC 935] | 379 | 348 | 492 |
| System (b) [Self-made LE discriminator] | 331 | 295 | 416 |

TRS= Coincidence time resolution of the BaF2 detector and the LYSO detectors;
TRL=deconvolved time resolution—fraction of the BaF2-XP2020Q detector deducted ;
TR=the expected coincidence time resolution of two identical LYSO detectors .

The results below were all measured by test system (b).

Figure 11 shows the coincidence time resolution (FWHM) of the 5×5 LYSO array detectors and the BaF$_2$ detector (TR$_S$), the time resolution (FWHM) of the 5×5 LYSO array detectors (TR$_L$), and the expected coincidence time resolution (FWHM) of two identical 5×5 LYSO array detectors, whose average value was 415.2 ps.

Table 2 shows the peak position of the time spectra of the crystals in the array, withthe calculated relative difference.

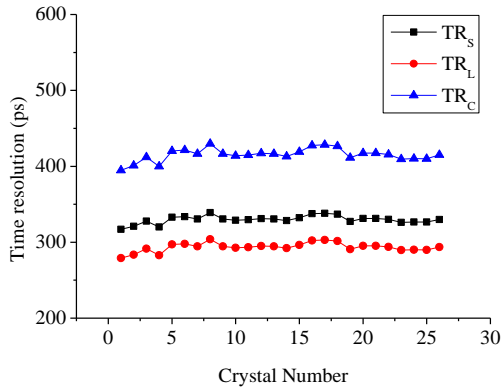

Figure 11 Time resolution of array detectors; TRS—Coincidence time resolution of the BaF2 detector and the LYSO detectors; TRL—Time resolution of the LYSO detectors; TRC—Coincidence time resolution of two identical LYSO detectors .

Table.2 The position of the time spectrum peaks for the 25 crystals in the array (1ch=25.1ps)

| Crystal number | Time spectrum peaks (channel) | Relative difference of the peak (ps) |
|---|---|---|
| 1 | 704.36 | 144 |
| 2 | 703.9 | 132 |
| 3 | 704.26 | 141 |
| 4 | 704.92 | 158 |
| 5 | 706.04 | 186 |
| 6 | 704.07 | 136 |
| 7 | 703.37 | 119 |
| 8 | 705.45 | 171 |
| 9 | 706.21 | 190 |
| 10 | 706.82 | 205 |
| 11 | 700.92 | 57 |
| 12 | 700.7 | 52 |
| 13 | 702.37 | 94 |
| 14 | 702.7 | 102 |
| 15 | 704.78 | 154 |
| 16 | 701.89 | 82 |
| 17 | 701.37 | 69 |
| 18 | 702.53 | 98 |
| 19 | 702.83 | 105 |
| 20 | 704.74 | 153 |
| 21 | 700.64 | 50 |
| 22 | 698.64 | 0 |
| 23 | 700.05 | 35 |
| 24 | 700.14 | 38 |
| 25 | 701.84 | 80 |

The timing performance of the LYSO detectors was tested under different trigger threshold voltages for the LE discriminator. The average expected coincidence time resolution of two identical LYSO detectors with the change of the threshold is shown in Figure 12.

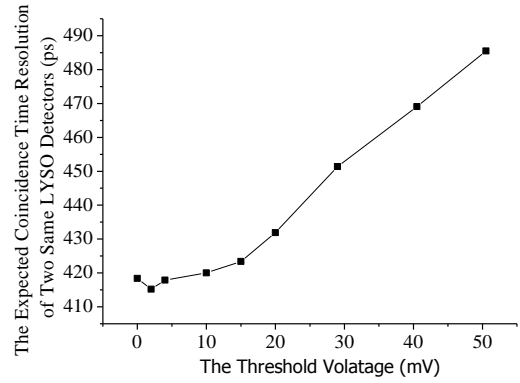

Figure 12 Mean coincidence time resolution for two identical LYSO detectors for the 5×5 array

## 4. Discussion

The factors influencing timing resolution in a TOF-PET detector mainly include the light decay curve of the crystal, the yield of the light, the size of the crystal, the light collection method, the PMT and the timing discriminator [10~11]. As the timing resolution of a TOF-PET is required to be very high, every detail of the experiment is important.

In this work, the rise time of the LYSO signal tested by the oscilloscope is 1.7 ns, close to the result (1.8 ns) measured by the single photon method, which means that the H8500 PMT is very fast, and has almost no effect on the quality of the scintillator. The variation in the peak position of the time spectra shown in Table 2 includes the contribution of the crystals and the PMT of the 16mm×16mm center area. To a certain extent, it reflects that the transit time spread of the PMT is relatively small, as the biggest variation is 190ps. This is because that the transit time spread of the H8500 PMT is only 400 ps for the whole effective area [12], comparable to other fast timing PMTs. The fast rise time and the small transit time spread are the important factors that give the H8500 PMT good time performance.

Using CFD as the timing discriminator, the coincidence time resolution (3.2mm×3.2mm×25mm LYSO, LYSO to LYSO, CFD) is FWHM 492 ps, obviously better than the result obtained by T. Moriya (FWHM 505 ps, 2.9mm×2.9mm×20mm LYSO, LYSO to BaF₂, CFD)[7], but worse than the result achieved by Chang Lyong Kim (477 ps, 4.2mm×4.2mm×30mm LYSO, LYSO to LYSO, LE discriminator)[4], as longer crystals give worse time resolution[12]. One reason might be the size of the crystal used by T. Moriya was slimmer. Another reason might be the array structure used in this work and Chang Lyong Kim's work is better for timing resolution. The latter both used air-gaps for a large area of the array. Slimmer crystals mean better position resolution. Longer crystals mean higher detection efficiency. The cross section of the TOF-PET spreads from 2.9 mm ×2.9mm to 6.75 mm ×6.75mm )[ 7,11~14] . The balance of time resolution, position resolution and detection efficiency is a complex problem to consider for commercial TOF-PET equipment.

As can be seen from Table 1, the time resolution of the LYSO detector is 53 ps better when "the inverter + CFD ORTEC 935" is changed to the LE discriminator as timing discriminator. One reason for this is that LE discriminator is more suitable for the LYSO than CFD [8]. Time walk because of the variation of the amplitude of the signal is the main problem for LE discriminators. However, the rise time of LYSO scintillator is very quick (1.8 ns). Furthermore, only the 511 keV peak area is used for TOF-PET. The light yield of LYSO crystal standing on the PMT was 3226 p.e/MeV, which made the variation of the signal amplitude very small (good energy resolution). Besides, the threshold voltage of the LE discriminator was very low, only 1% (in this paper 2 mV) of the signal amplitude. The time walk decreases when the threshold is lowered, as illustrated in Figure.13. These three reasons together made the time walk very small. The timing variation caused by noise for the LE discriminator is smaller than that for CFD. Besides, CFD needs one signal to be delayed and another signal to be inverted. Usually, commercial CFDs are designed for negative input signals. For the H8500, however, the anode signal is not convenient for timing. The dynode signal has to be inverted to fit the CFD. Every processing step will cause some degradation in the time performance of the signal. The reasons above make the LE discriminator better than CFD for our purposes.

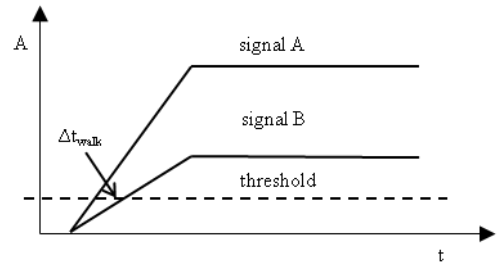

Figure.13 illustration of time walk for LE discriminator

It can be concluded from Figure 12 that the time resolution of the LYSO array detectors becomes better with the decrease of the LE discriminator threshold of the, and the best time resolution can be obtained with the threshold voltage at about 1% (in this paper 2 mV) of the signal amplitude. When the threshold approaches 0 mV, the time resolution becomes worse. This is because the noise triggering becomes dominant.

The average coincidence time resolution of two identical LYSO arrays was FWHM 415.2 ps, good enough for TOF-PET. It can be expected that the 5×5 LYSO array can be expanded to a 15×15 array to cover the whole effective area of the H8500 PMT without degradation of the time performance.

### Acknowledgements

This work was Supported by National Natural Science Foundation of China (10775149, 10805049). One of the authors (Cheng Fengfeng) would like to acknowledge support by "the Fundamental Research Funds for the Central Universities" (2013NT10).